\documentclass[lettersize,journal]{IEEEtran}
\usepackage{amsmath,amsfonts}
\usepackage{algorithmic}
\usepackage{algorithm}
\usepackage{array}
\usepackage[caption=false,font=normalsize,labelfont=sf,textfont=sf]{subfig}
\usepackage{textcomp}
\usepackage{stfloats}
\usepackage{url}
\usepackage{verbatim}
\usepackage{graphicx}
\usepackage{cite}
\begin{document}

\title{Drift-Aware Adaptive Semantic Codec Update Strategy for Dynamic Semantic Communications}

\author{{Bizhu Wang, Manqi Li, Shujun Han, IEEE member, Mengying Sun, IEEE member,\\ Xiaodong Xu, IEEE Senior Member}
\thanks{Bizhu Wang, Manqi Li, Shujun Han, Mengying Sun and Xiaodong Xu are with the State Key Laboratory of Networking and Switching Technology, Beijing University of Posts and Telecommunications (BUPT), Beijing, China. (Corresponding author: Xiaodong Xu)}}

\markboth{Journal of \LaTeX\ Class Files,~Vol.~14, No.~8, August~2021}%
{Shell \MakeLowercase{\textit{et al.}}: A Sample Article Using IEEEtran.cls for IEEE Journals}


\maketitle

\begin{abstract}
Semantic communication reduces transmission overhead by conveying task-relevant semantic features, but its performance depends on the consistency between training and target-domain data distributions. In dynamic environments, distribution drift can make offline-trained semantic codec models stale and degrade transmission performance. Frequent updates alleviate this degradation but increase communication, computation, and queueing costs, whereas infrequent updates delay model adaptation. To address this tradeoff, we propose a drift-aware adaptive update strategy for dynamic semantic communication systems. Drift intensity is quantified in the semantic feature space, and Drift-Aware Age of Model (DA-AoM) is introduced to jointly capture drift-induced performance degradation, drift-dependent update service delay, and conventional update latency. Closed-form average DA-AoM expressions are derived for edge-layer M/M/1 and cloud-layer D/M/1 update scenarios, based on which the update frequency is optimized under the queueing-stability constraint. Simulations show that DA-AoM captures the joint effects of drift intensity, computing rate, bandwidth, and update frequency on model freshness beyond conventional Age of Information (AoI). The proposed strategy reduces average DA-AoM by 9.2\%--42.4\% compared with representative baselines under the same resource constraints.
\end{abstract}

\begin{IEEEkeywords}
semantic communication, data distribution drift, drift-aware age of model, model update strategy, resource constraints
\end{IEEEkeywords}

\section{Introduction}

\IEEEPARstart{T}{he} evolution of 6G networks and emerging applications, such as extended reality and autonomous driving, has increased bandwidth demands on wireless networks \cite{ref1,ref2,ref3}. Under limited spectrum and edge computing resources, conventional bit-level communication paradigms struggle to meet these efficiency requirements \cite{ref2,ref4}. Semantic communication (SemCom) \cite{ref4,ref5,ref6} addresses this challenge by prioritizing the delivery of information meaning and task objectives rather than exact bit recovery \cite{ref7,ref8,ref9,ref10}. A typical SemCom system trains semantic codec models offline on resource-rich cloud or edge servers and deploys lightweight versions on end devices for real-time inference \cite{ref11,ref12,ref13}.\par

In practical deployments, data distributions shift with factors such as weather, task demands, and user behavior \cite{ref14,ref15}. For semantic codecs, such distribution drift makes offline-trained models stale, reducing target-domain semantic transmission quality and weakening SemCom system reliability \cite{ref8,ref16}. Timely updates can mitigate this degradation, but maintaining model freshness introduces resource constraints. Transmitting raw target-domain data for model updates consumes scarce wireless bandwidth, while executing the update process imposes substantial computing load \cite{ref11,ref12,ref13,ref17}. A high update frequency can cause congestion and service delay \cite{ref17,ref18}, whereas a low update frequency prevents the codec from adapting promptly to distribution shifts, leading to prolonged performance degradation. Therefore, adaptive semantic codec updating under resource constraints requires a freshness metric that accounts for drift-induced model degradation.\par


Existing status update systems mainly rely on Age of Information (AoI) \cite{ref19,ref20,ref21}, defined as the elapsed time since the last received update was generated, and several related metrics including Age of Processing (AoP) \cite{ref22}, Age of Task (AoT) \cite{ref23}, Age of Incorrect Information (AoII) \cite{ref24}, and Age of Model (AoM) \cite{ref25}. Although these metrics extend freshness measurement to computation, task completion, and information utility, they remain primarily latency-driven and do not capture performance degradation caused by source-target distribution shifts.\par

Existing model update strategies mainly adopt fixed-period, threshold-triggered, or state-adaptive policies. Fixed-period updating cannot respond to changing distribution drift \cite{ref19,ref21,ref25}, while threshold-triggered schemes
depend on predefined thresholds and may lead to abrupt update decisions \cite{ref26,ref27,ref28}. Adaptive strategies can adjust the update frequency according to system states, but typically rely on heuristic mappings and do not explicitly characterize the joint effects of drift-induced model degradation and update queueing delay \cite{ref17,ref18,ref29}. Therefore, an update strategy that jointly accounts for semantic model freshness and resource-constrained update delay is still needed.


To address this gap, this paper first introduces Drift-Aware Age of Model (DA-AoM) to characterize semantic codec freshness under continuous data distribution drift. Based on the resulting drift-aware freshness model, we further formulate the semantic codec update-frequency optimization problem under communication and computation constraints. The proposed framework provides an end-to-end analytical path from drift quantification and freshness modeling to update-frequency optimization. The main contributions are summarized as follows:\par

\begin{itemize}
    \item We propose a drift-aware adaptive semantic codec update strategy for dynamic SemCom systems, which integrates data distribution drift quantification, DA-AoM-based model freshness evaluation, and update-frequency adaptation under communication and computation constraints.
    
    \item We characterize data distribution drift in the semantic feature space to provide the drift state required for subsequent freshness modeling. Source- and target-domain features extracted by SwinJSCC are represented using multidimensional semantic distances, followed by drift-rank-based normalization and mild polarization adjustment to obtain the drift intensity and establish its association with the model performance degradation rate.
    
    \item We introduce DA-AoM as a drift-aware model freshness metric by incorporating the performance degradation rate and drift-dependent computational load into the age evolution. Based on queueing analysis, closed-form expressions of the average DA-AoM are derived for edge-layer M/M/1 random-update and cloud-layer D/M/1 periodic-update scenarios, jointly characterizing drift-induced model degradation and model update delay.
    
    \item We design a DA-AoM-driven adaptive semantic codec update strategy under the queueing-stability constraint. By normalizing the actual update frequency according to the drift-dependent service capability, we characterize the non-monotonic relationship between DA-AoM and update frequency, determine the optimal normalized update frequencies for the M/M/1 and D/M/1 scenarios, and map them to drift-adaptive actual update frequencies.
    
    \item Simulation results show that DA-AoM captures the effects of drift intensity, computing rate, network bandwidth, and update frequency on model freshness beyond conventional AoI. The proposed update strategy reduces the average DA-AoM by 9.2\%–42.4\% compared with proportional, threshold-triggered, square-root, fixed low-frequency, and fixed high-frequency strategies.
\end{itemize}

The rest of this paper is organized as follows. In Section~\ref{section3}, the association between data distribution drift measurement and the model performance degradation rate is constructed. The DA-AoM model is developed in Section~\ref{section4}, where its closed-form expressions at the edge and cloud layers are derived. Section~\ref{section5} proposes a DA-AoM-based adaptive optimal model update strategy. The simulation results are presented in Section~\ref{section6} to verify the effectiveness of the DA-AoM metric and analyze the performance advantages of the proposed model update strategy over baseline strategies. Finally, Section~\ref{section7} concludes the paper.

\section{Data Distribution Drift Measurement and Performance Degradation Modeling}
\label{section3}
In SemCom systems, the data distribution difference between the source and target domains is a key factor affecting the performance of semantic codec models \cite{ref30,ref31,ref32}, such as transmission performance measured by peak signal-to-noise ratio (PSNR). To find the optimal update frequency under resource constraints in data distribution drift scenarios, it is first necessary to accurately measure the current data distribution drift and then evaluate the performance loss rate of the semantic codec models under the current drift intensity. This provides a basis for subsequent decisions such as update frequency selection. To achieve this goal, this section proposes a data distribution drift measurement method and establishes the association expression between data distribution drift and the performance degradation rate of semantic codec models.

\subsection{Data Distribution Drift Intensity Quantification Method}
\label{section3.1}
\subsubsection{Feature Extraction Network Based on SwinJSCC and Attention Enhancement}
\label{section3.1.1}
\begin{figure*}[t]
    \centering
    \includegraphics[width=0.95\textwidth]{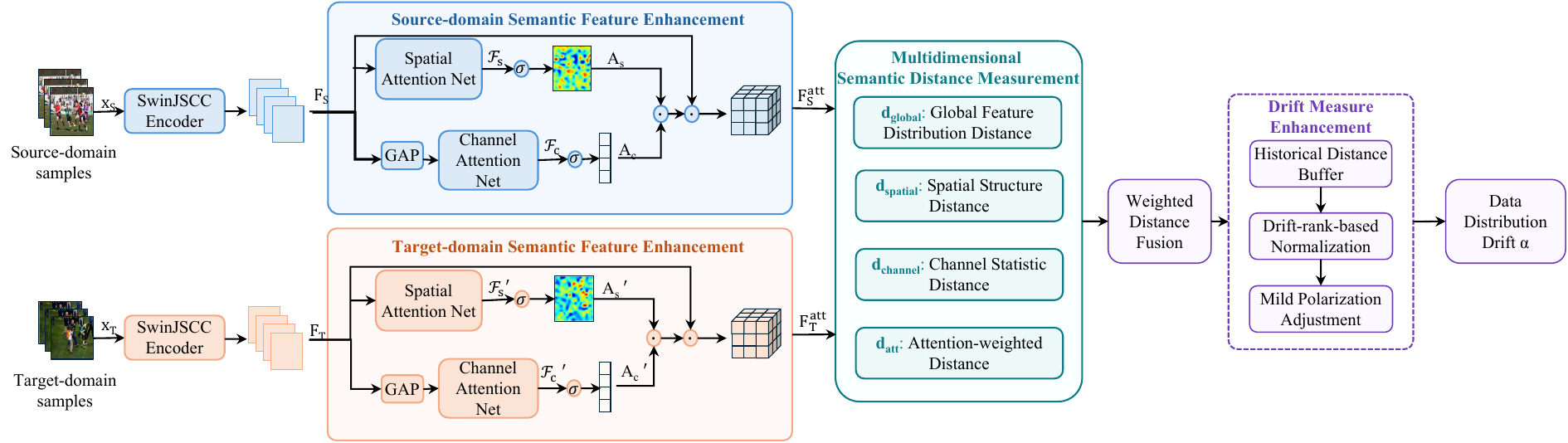}
    \caption{Data distribution drift intensity quantification framework.}
    \label{fig:framework}
    \vspace{-5mm}
\end{figure*}

As shown in Fig.~\ref{fig:framework}, SwinJSCC-Large \cite{ref33} is adopted to extract latent semantic features from the source- and target-domain samples. For an input image $\mathbf{x}\in\mathbb{R}^{H\times W\times 3}$, the encoder ${E}_{\theta}$ produces $\mathbf{F}={E}_{\theta}(\mathbf{x}) \in\mathbb{R}^{h\times w\times C}$. To emphasize semantically informative spatial regions and feature channels, spatial and channel attention are applied in parallel, yielding the attention-enhanced feature
\begin{equation}
\mathbf{F}^{\mathrm{att}}
=
\mathbf{F}
\odot
\sigma\!\left(\mathcal{F}_{s}(\mathbf{F})\right)
\odot
\sigma\!\left(
\mathcal{F}_{c}(\mathrm{GAP}(\mathbf{F}))
\right).
\label{eq:att_feature}
\end{equation}
The resulting source- and target-domain representations are denoted by $\mathbf{F}^{\mathrm{att}}_{S}$ and $\mathbf{F}^{\mathrm{att}}_{T}$, respectively.

\subsubsection{Multidimensional Semantic Distance Measurement}

Based on the attention-enhanced source- and target-domain features
$\mathbf{F}^{\mathrm{att}}_{S}$ and
$\mathbf{F}^{\mathrm{att}}_{T}$, four complementary semantic
distance metrics are used to characterize distribution differences
from global, spatial, channel, and attention-aware perspectives.
Specifically, $d_{\mathrm{global}}$ measures the domain-level
feature deviation, $d_{\mathrm{spatial}}$ captures changes in
spatial structure, $d_{\mathrm{channel}}$ characterizes
channel-wise statistical shifts, and $d_{\mathrm{att}}$ emphasizes
feature differences in semantically salient regions. These metrics
are defined as
\begin{equation}
\begin{aligned}
d_{\mathrm{global}}
&=
\left\|
\mathrm{GAP}\!\left(\mathbf{F}^{\mathrm{att}}_{S}\right)
-
\mathrm{GAP}\!\left(\mathbf{F}^{\mathrm{att}}_{T}\right)
\right\|_{2},
\\
d_{\mathrm{spatial}}
&=
\frac{1}{hw}
\sum_{p=1}^{h}
\sum_{q=1}^{w}
\left\|
\mathbf{F}^{\mathrm{att}}_{S}(p,q)
-
\mathbf{F}^{\mathrm{att}}_{T}(p,q)
\right\|_{2},
\\
d_{\mathrm{channel}}
&=
\frac{1}{C}
\sum_{c=1}^{C}
\left\|
\mathrm{avg}\!\left(\mathbf{F}^{c}_{S}\right)
-
\mathrm{avg}\!\left(\mathbf{F}^{c}_{T}\right)
\right\|_{2},
\\
d_{\mathrm{att}}
&=
\frac{1}{N}
\sum_{i=1}^{N}
\left\|
\left(
\mathbf{F}^{\mathrm{att}(i)}_{S}
-
\mathbf{F}^{\mathrm{att}(i)}_{T}
\right)
\odot
\frac{
\mathbf{A}^{(i)}_{s}
+
\mathbf{A}^{(i)}_{c}
}{2}
\right\|_{F}.
\end{aligned}
\label{eq:semantic_distances}
\end{equation}

The four distances provide complementary descriptions of semantic
distribution shift and are linearly fused to obtain the raw drift
metric
\begin{equation}
d_{\mathrm{raw}}
=
w_{1}d_{\mathrm{global}}
+
w_{2}d_{\mathrm{spatial}}
+
w_{3}d_{\mathrm{channel}}
+
w_{4}d_{\mathrm{att}},
\label{eq:raw_drift}
\end{equation}
where $w_{1}$--$w_{4}$ denote the corresponding fusion weights.

These distance components constitute an image-codec instantiation
of the proposed drift quantification method rather than a strict
architecture-independent requirement of DA-AoM.
While $d_{\mathrm{global}}$ can be constructed directly from latent
semantic representations, $d_{\mathrm{spatial}}$,
$d_{\mathrm{channel}}$, and $d_{\mathrm{att}}$ exploit the spatial,
channel, and attention structures of the image codec.
For other semantic modalities, these components can be replaced
with representation-specific semantic distances, while the
subsequent drift normalization, enhancement, and DA-AoM modeling
remain unchanged.
\subsubsection{Enhanced Data Distribution Drift Measurement}

The multidimensional semantic distances defined in
\eqref{eq:raw_drift} provide an initial measure of the distribution
difference between the source and target domains. However, the raw
drift metric $d_{\mathrm{raw}}$ has an unbounded numerical range,
which makes a universal threshold difficult to define across
different distribution states \cite{ref34,ref35}. Moreover, small
semantic differences may correspond to very similar L2 distances in
the feature space, reducing the sensitivity of $d_{\mathrm{raw}}$.
To address these limitations, the raw drift metric is further
processed through two steps: drift rank normalization and mild
polarization adjustment.

\textit{(1) Drift rank normalization:}
For each distance component
$d_i\in\{d_{\mathrm{global}},d_{\mathrm{spatial}},
d_{\mathrm{channel}},d_{\mathrm{att}}\}$, a historical distance
buffer is maintained to estimate its mean $\mu_i$ and standard
deviation $\sigma_i$. The current distance is first standardized
according to its relative position in the historical distribution,
and then mapped to the interval $[0,1]$ through a Sigmoid function as
\begin{equation}
    z_i=\frac{d_i-\mu_i}{\sigma_i+\epsilon},
    \qquad
    \bar{d}_i=\frac{1}{1+\exp(-kz_i)},
    \label{eq:rank_normalization}
\end{equation}
where $\epsilon$ avoids numerical instability and $k$ controls the
sensitivity of the mapping. The normalized distance components are
then combined through weighted fusion to obtain
$\widehat{d}=\sum_{i=1}^{4}w_i\bar{d}_i$.

\textit{(2) Mild polarization adjustment:}
Although drift rank normalization improves the comparability of the
distance components, differences in the small and large drift regions
may still be insufficiently distinguished. Therefore, a piecewise
nonlinear mapping is applied to $\widehat{d}$. Two thresholds,
$\tau_{\mathrm{low}}$ and $\tau_{\mathrm{high}}$, are introduced to
divide the normalized drift into three regions. For
$\widehat{d}<\tau_{\mathrm{low}}$, a power function with
$\gamma_{\mathrm{low}}>1$ suppresses minor feature fluctuations and
reflects the robustness of the semantic codec to slight distribution
changes. For $\widehat{d}>\tau_{\mathrm{high}}$, a power function
with $\gamma_{\mathrm{high}}<1$ enhances the distinction among severe
distribution shifts. The intermediate region remains unchanged to
preserve a smooth variation of the drift metric.

Through the above two steps, the raw drift metric
$d_{\mathrm{raw}}$ is transformed into the enhanced data distribution
drift intensity $\alpha$, which is defined as
\begin{equation}
    \alpha=
    \begin{cases}
        \widehat{d}^{\,\gamma_{\mathrm{low}}},
        & \widehat{d}<\tau_{\mathrm{low}},\\
        1-\left(1-\widehat{d}\right)^{\gamma_{\mathrm{high}}},
        & \widehat{d}>\tau_{\mathrm{high}},\\
        \widehat{d},
        & \text{otherwise}.
    \end{cases}
    \label{eq:alpha}
\end{equation}

\begin{figure}[!t]
    \centering
    \includegraphics[width=0.9\linewidth]{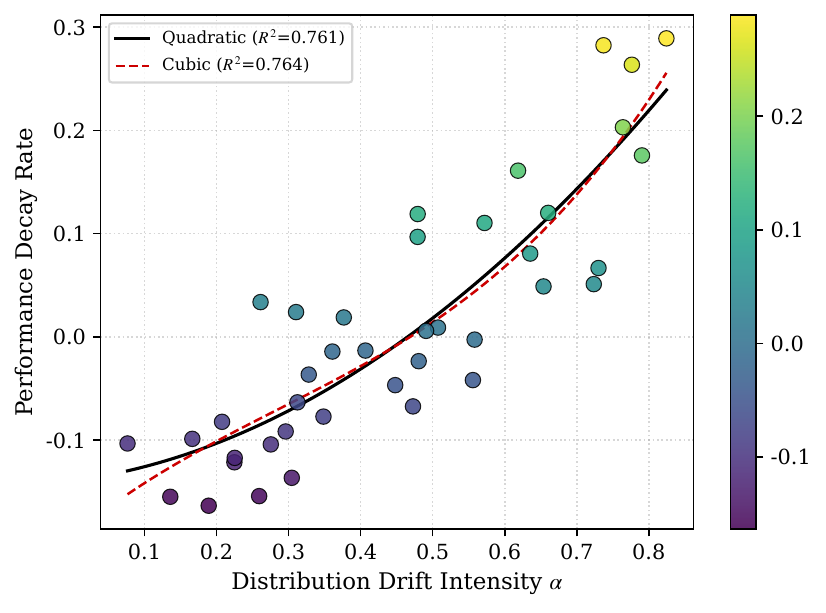}
    \caption{Association between enhanced data distribution drift
    $\alpha$ and the performance degradation rate of semantic codec
    models.}
    \label{fig3:alpha}
    \vspace{-5mm}
\end{figure}

As shown in Fig.~\ref{fig3:alpha}, the enhanced drift intensity
$\alpha$ exhibits a clearer nonlinear association with the model
performance degradation rate. Compared with the raw drift metric,
the coefficient of determination of the cubic fitting increases from
$R^2=0.569$ to $R^2=0.764$. This result indicates that the proposed
normalization and polarization procedure improves the ability of the
drift metric to characterize performance degradation caused by data
distribution changes.
\subsection{Association Model of the Performance Degradation Rate of Semantic Codec Models}
\label{section3.1.2}
Based on the characterization of the data distribution drift intensity $\alpha$ in Section~\ref{section3.1}, this section first calculates the performance degradation rate of the semantic codec models under different drift intensities by using transmission performance metrics such as PSNR. Then, a nonlinear polynomial regression method is used to establish the association expression $u(\alpha)$ between the data distribution drift intensity $\alpha$ and the model performance degradation rate.

\subsubsection{Calculation of the Model Performance Degradation Rate}

After an image $I$ is input into the semantic codec model, it is mapped into semantic features by the semantic encoder and transmitted through the channel. At the receiver, the received signal is decoded to reconstruct the image $\widehat{I}$. The performance metric between the original image $I$ and the reconstructed image $\widehat{I}$, such as PSNR or other semantic distortion metrics, is calculated and averaged to obtain the source-domain baseline performance $P_S$ and the measured target-domain performance $P_T^\alpha$ under the current drift intensity $\alpha$. Since the target-domain data distribution differs from the source-domain distribution, the relative loss of $P_T^\alpha$ with respect to $P_S$ is calculated as the model performance degradation rate $u_\alpha$, given by

\begin{equation}
    u_\alpha=\frac{P_S-P_T^\alpha}{P_S}.
    \label{eq:u}
\end{equation}

Although image reconstruction is considered in this work, $P_S$ and $P_T^\alpha$ can generally represent the source- and target-domain performance of other semantic tasks, with the metric direction normalized such that a larger value indicates better performance. For example, classification accuracy or semantic/task similarity can replace reconstruction quality in task-oriented, text, or speech semantic communication. Extending DA-AoM to another task therefore requires redefining the task-specific performance measure and fitting the corresponding $u(\alpha)$, while the subsequent DA-AoM formulation and update-frequency analysis remain unchanged.

\subsubsection{Fitting of the Model Performance Degradation Rate Association Model}

Based on the model performance degradation rate $u_\alpha$, each target-domain subdomain is divided into a training set and a test set. On the training set, the data distribution drift intensity $\alpha$ of each subdomain relative to the source domain and the actual performance degradation $u_\alpha$ are calculated. Polynomial regression is then used to fit the association function $u(\alpha)$ between the drift intensity $\alpha$ and the model performance degradation rate, and the fitted function is validated on the test set of each subdomain.

Based on the analysis of data distribution drift in Section~\ref{section3.1}, the model is robust to slight environmental perturbations in the low-drift region, and its performance degrades slowly. In some samples, the performance even remains stable or slightly improves. However, once the drift exceeds a critical threshold, model performance shows an accelerated degradation trend \cite{ref36}. Based on this property, this paper compares quadratic and cubic polynomial fitting. As shown in Figure~\ref{fig3:alpha}, the cubic polynomial achieves a goodness of fit of $R^2=0.764$ on the training set, and can more accurately capture the nonlinear process between the drift intensity $\alpha$ and the model performance degradation rate $u(\alpha)$. The association function is expressed as

\begin{equation}
    u(\alpha)=c_3\alpha^3+c_2\alpha^2+c_1\alpha+c_0,
    \label{eq:u_fitting}
\end{equation}

where the polynomial coefficients $c_0$, $c_1$, $c_2$, and $c_3$ can be determined by least-squares regression. Compared with the fitting curve obtained from the original data distribution drift, the cubic polynomial fitted using the enhanced drift intensity deviates significantly from the quadratic curve in the high-drift region. This indicates that the enhanced drift metric not only improves $R^2$ from 0.569 to 0.764, but also enables the cubic term to capture additional nonlinear information, verifying the rationality of the selected model complexity.

\begin{figure}[!t]
    \centering
    \includegraphics[width=0.9\linewidth]{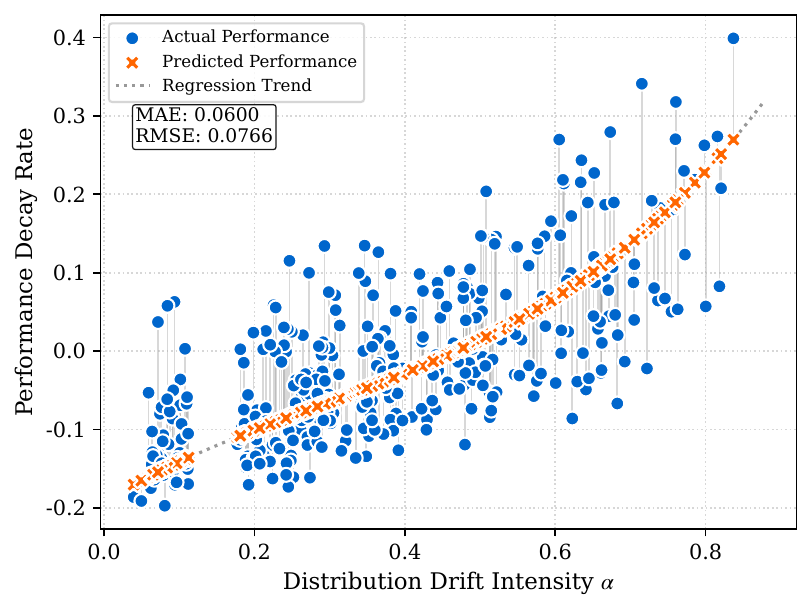}
    \caption{Validation of the association function $u(\alpha)$ on the test set: comparison between actual degradation and predicted degradation.}
    \label{fig:test}
    \vspace{-5mm}
\end{figure}

As shown in Figure~\ref{fig:test}, the fitted function is further validated on the test set. Although semantic-L2-distance-based clustering is not performed on the test set and the inherent content variance of individual samples is retained, the association model between drift intensity and model performance degradation still achieves a goodness of fit of $R^2=0.51$ and an RMSE of only 0.07. This result indicates that the predicted curve can capture the overall degradation trend with acceptable error, proving that the proposed enhanced drift intensity $\alpha$ can effectively characterize semantic distribution shifts in the feature space. The constructed association function $u(\alpha)$ has good generalization ability and can provide a trend prediction with controllable bias for model performance degradation in unseen scenarios.

\section{DA-AoM Theoretical Framework and Mathematical Model}
\label{section4}
 To cope with the negative impact of data distribution drift on semantic codec performance described above, the model on the device side must be updated in time \cite{ref37}. However, model updating often relies on data transmission from devices to edge or cloud servers. Under limited bandwidth and computing resources, overly frequent update strategies may cause network congestion or competition for computing resources. To address this challenge, AoI can be used to quantify the temporal freshness of data at the receiver and guide model updates. On this basis, AoM was proposed to measure model freshness while further considering the impact of training data scale on model performance, that is, the utility of freshness, thereby guiding model updates. 

 Inspired by AoM proposed in \cite{ref25}, this section incorporates data distribution drift intensity into AoI and proposes the DA-AoM. It investigates how data distribution drift intensity affects the freshness of semantic codec models, namely model freshness, under resource constraints, laying the foundation for the model update strategy in Section~\ref{section5}.

 This section first formalizes the model performance degradation rate $u(\alpha)$ according to the definition of AoM. Considering the network resource constraints in SemCom systems, it further establishes a model update computational load and delay measurement model that includes a theoretical drift-aware dynamic load mechanism and a measured baseline fixed load mechanism, in order to describe the constraining relationship of data distribution drift on system resources. On this basis, this section derives the average DA-AoM expressions for model updates performed at the edge-layer and cloud-layer servers, respectively.

\subsection{Basic Framework and Mathematical Definition of DA-AoM}
\label{section4.1}

To quantify the impact of data distribution drift on the freshness of semantic codec models, this section proposes DA-AoM. It describes the cumulative model performance loss of the system at any time $t$, and is jointly determined by temporal freshness and the model performance degradation rate. Unlike the original AoM in \cite{ref25}, where $u(\cdot)$ is only related to the number of training categories, this paper relates $u(\cdot)$ to data distribution drift. Based on the cubic polynomial fitted in Section III, a nonnegative
lower bound $\epsilon$ is introduced to avoid numerical anomalies,
and the degradation function is expressed as

\begin{equation}
    u(\alpha) = \max\left( \epsilon,c_{3}\alpha^{3} + c_{2}\alpha^{2} + c_{1}\alpha + c_{0} \right)
    \label{u_correct}
\end{equation}

\begin{figure}[!t]
    \centering
    \includegraphics[width=0.9\linewidth]{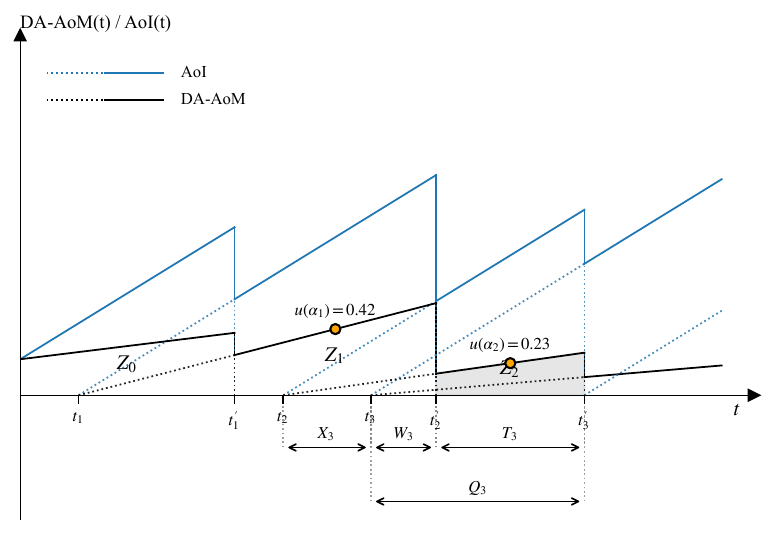}
    \caption{Evolution of DA-AoM and AoI within an update cycle.}
    \label{fig:daaom evolution}
    \vspace{-5mm}
\end{figure}

This section derives the average DA-AoM state in SemCom systems by drawing on the method in \cite{ref25}. Figure~\ref{fig:daaom evolution} illustrates an example of how DA-AoM and AoI change over time $t$ as the semantic codec model is continuously updated. Assume that observation begins at $t=0$, at which time the model update queue is empty and the initial value is $\text{DA-AoM}_{(t_0)}$. The $i$-th model update request is generated at time $t_i$, with subsequent updates occurring sequentially at $t_2,t_3,\cdots,t_n$. During intervals when no new model updates are completed, the value of DA-AoM increases linearly over time and is reset to a smaller value upon completion of an update. The update generated at time $t_i$ is completed at time $t_i'$.

After the $i$-th update request is processed at $t_i'$, the DA-AoM value $DA-AoM(t)$ is reset to $DA-AoM\left(t_i^\prime\right)=Q_i\left(u\left(\alpha\right)\right)$, where $Q_i$ is the total duration of the model update process, which is the sum of the time spent waiting in the queue and the time spent in service. $u(\alpha)$ is the model performance degradation rate. However, at $t_i'$, the AoI value $AoI(t)$ is reset to $AoI\left(t_i^\prime\right)=Q_i\ast{u_{AoI}}$, and $u_{AoI}$ is a fixed value, without considering update differences. Because the model usually has a certain generalization capability, it typically satisfies $u(\alpha)<1$. Therefore, the DA-AoM curve in the figure remains below the AoI curve, indicating that the performance degradation rate at the semantic level is lower than the staleness growth rate at the purely temporal level.


Following the cycle-based model freshness analysis in AoM \cite{ref25}, we characterize the evolution of DA-AoM according to the distribution state associated with each model update cycle. Concept-drift studies commonly model non-stationary data streams as a sequence of locally stationary contexts, while the underlying distribution may change between contexts \cite{ref26,ref35}. Accordingly, for tractable per-cycle analysis, we assume that the target-domain distribution is approximately stationary within a single model update cycle. Thus, the drift intensity $\alpha_i$ and corresponding performance degradation rate $u(\alpha_i)$ are treated as locally constant within the $i$-th cycle, while $\alpha_i$ may vary across different cycles as the target-domain environment evolves. This assumption does not require the data distribution drift to remain stationary over the entire system operation.

After the $i$-th model update is completed at time $t_i'$, the system uses the model update generated at time $t_i$ until the next update is completed. Under the current cycle-level drift state $\alpha_i$, the instantaneous DA-AoM is defined as
\begin{equation}
    \Delta_{\text{DA-AoM}}(t) = u(\alpha) \times \left( t - t_{i} \right), t_i' \le t < t_{i+1}'.
\label{eq:daaom instantaneous}
\end{equation}

After the instantaneous DA-AoM is given, this section further derives the average DA-AoM state using a geometric decomposition method similar to that in \cite{ref25}, so as to better represent the system problem. To calculate the average DA-AoM for the observation period from $t=0$ to $t=T$, this section computes the total area under the waveform and divides it by the total time $T$. The total area can be decomposed into the sum of a series of non-overlapping geometric regions $Z_i$, as shown by the shaded area $Z_2$ in Figure~\ref{fig:daaom evolution}.

Consider the times $t_{i-1}'$ and $t_i'$ when two consecutive updates are completed. During this period, the  model always infers based on the results of the $i-1$th update, with a corresponding data distribution drift intensity of $\alpha_{i-1}$. The DA-AoM evolution slope in this interval is constant $u(\alpha_{i-1})$. The integral within each cycle $Z_i$ can be represented as the area of a right-angled trapezoid with height $T_i$. The area of $Z_i$ can be calculated as the difference between the areas of two isosceles triangles formed by the points $t_{i-1}$ and $t_i'$, and $t_{i-1}$ and $t_{i-1}'$. Defining $X_i=t_i-{t_{i-1}}$ as the time interval between the arrival of the $i-1$th and $i$th update requests, the area of $Z_i$ can be expressed as

\begin{equation}
    \begin{array}{r} Z_{i} = \int_{t_{i - 1}^{'}}^{t_{i}^{'}} u\left( \alpha_{i - 1} \right) \cdot \left( t - t_{i - 1} \right)dt \\ = \frac{1}{2}u\left( \alpha_{i - 1} \right)\left[ \left( X_{i} + W_{i} + T_{i} \right)^{2} - (W_{i} + T_{i})^{2} \right] \end{array}.
    \label{eq:Z_i}
\end{equation}

Assume that the system enters steady state and that the update process is stationary and ergodic. At observation time $T\rightarrow \infty$, the influence of boundary regions $Z_0$ and $Z_n$ can be ignored. The average DA-AoM can be expressed as the sum of all cycle areas divided by the total time

\begin{equation}
    {\overline{\Delta}}_{DA\text{-}AoM} = \lim_{T \rightarrow \infty}\frac{\sum_{i}^{} Z_{i}}{T} = \frac{E\left[ Z_{i} \right]}{E\left[ X_{i} + Q_{i} \right]},
    \label{eq:average daaom}
\end{equation}

where $E[\cdot]$ denotes the expected value. Considering the steady-state rate of the model update process, under the update frequency $\lambda$, the average DA-AoM can be expressed as

\begin{equation}
    {\overline{\Delta}}_{\text{DA-AoM}} = \frac{1}{2}\lambda u\left( \overline{\alpha} \right)E\left[ \left( {X_{i}}^{2} + 2X_{i}Q_{i} \right) \right].
    \label{eq:average daaom_lam}
\end{equation}

\subsection{Model Update Computational Load and Delay Measurement Model}
\label{section4.2}
In SemCom systems, data distribution drift $\alpha$ not only affects the performance degradation rate of semantic codec model, but also further affects the queueing waiting time and computation processing time at the computation node by changing the effective model update load. This section divides model delay into two layers: network transmission delay outside the computation queueing system, and computation processing delay of model updating that enters the M/M/1 or D/M/1 queueing analysis.

\subsubsection{Network Transmission Delay}

Network transmission delay is mainly constrained by the data size of the semantic model update parameters and the current communication-link bandwidth. Let $D_\text{model}$ denote the data size of a single model update package and $R_\text{channel}$ denote the effective transmission rate between the computing node and the data source. The network transmission delay $T_\text{net}$ is expressed as $T_\text{net} = {D_\text{model}}/{R_\text{channel}}$.

This part of delay reflects the impact of the communication link on model completion time. This paper assumes that the channel state is relatively stable within a single update cycle. Therefore, $T_\text{net}$ can be treated as a constant term. If the system delay on the computing node side is $T_\text{sys}$, the end-to-end completion delay for a single model update can be expressed as $T_\text{update}=T_\text{sys}+T_\text{net}$. Here, $T_\text{sys}$ consists of the queueing wait time and computation service time on the computing node side, while $T_\text{net}$ represents the additional fixed communication transmission delay. The stability conditions of the subsequent queuing model are still determined solely by the computation service process, rather than by $T_\text{update}$.

\subsubsection{Computational Processing Delay: Mapping Based on Relative Load}

Unlike network transmission delay, computation processing delay is directly affected by data distribution drift $\alpha$. When the drift between the target-domain data distribution and source-domain training data distribution increases, the system usually requires a higher effective update computational load to make the updated semantic codec re-adapt to the target domain. Therefore, this paper uses a drift-aware relative computational load factor to characterize the impact of data distribution drift on model updating. 

To compare the performance difference between the theoretical system and the measured system, this paper models the computational load of model updating as a theoretical drift-aware load and a measured fixed load, respectively. The difference between them is that the former assumes the system can adjust the amount of computation required to complete one effective model update according to drift intensity, whereas the latter corresponds to the actual execution mode under the current fixed network-structure deployment.

For the theoretical drift-aware load mechanism, let $\Phi(\alpha)$ denote the computational load required to complete one effective model update when the drift intensity is $\alpha$. The floating-point operations required for each model update, measured in giga floating point operations (GFLOPs), are used as the computational-load metric. Specifically, for a given drift intensity, the system performs updates under different transmission rates or computational loads and measures the degradation rate after updating. If a certain load can reduce the post-update degradation rate to a state satisfying the preset performance requirement, this load is considered to complete one effective model update service.

The computational load required for model updating increases
approximately monotonically with the data distribution drift
intensity. Based on the measured computational loads, a quadratic
polynomial
$\Phi(\alpha)=a_m\alpha^2+b_m\alpha+c_m$
is adopted to characterize this relationship while avoiding
overfitting. The coefficients $a_m$, $b_m$, and $c_m$ can be
determined according to different tasks. Since the queueing model
requires a normalized service load representation, the relative
computational load is further defined as
$\phi(\alpha)=\Phi(\alpha)/\Phi_0$, where $\Phi_0$ denotes the
computational load of a complete baseline semantic codec model
update.

Combining the analysis of network transmission delay and computation processing delay, this paper divides the delay in the model update process into delay inside the computation queueing system and delay outside the computation queueing system. The delay inside the computation queueing system consists of the queueing waiting time and computation service time of the model update task, and is characterized by an M/M/1 or D/M/1 queue. The delay outside the computational queueing system is represented by network transmission delay $T_{net}$, which is incorporated as a fixed additive term into the end-to-end model update completion delay of model updates. Therefore, when the drift intensity is $\alpha$, the average service time for model update tasks is $E[S] = {\phi(\alpha)}/{\mu}$.

\subsection{DA-AoM at the Edge-Layer and Cloud-Layer Model Updaters}
\label{section4.3}

Following the queueing abstraction adopted in AoM \cite{ref25}, we consider two representative model-update arrival patterns to obtain analytically tractable closed-form expressions. For edge-layer updates triggered by random environmental changes or user behaviors, update requests are modeled as Poisson arrivals with i.i.d. service times, yielding an M/M/1 queue. For cloud-layer periodic updates, requests are modeled with deterministic inter-arrival intervals and exponential service times, yielding a D/M/1 queue. These models characterize representative random and periodic update processes rather than restricting DA-AoM to these two arrival patterns.

\subsubsection{Edge DA-AoM With Single Server-M/M/1}

Factors affecting DA-AoM include update frequency $\lambda$, data distribution drift intensity $\alpha$, and computing rate $\mu$. Data distribution drift affects both the computational load $\phi(\alpha)$ and the model performance degradation rate $u(\alpha)$. In the M/M/1 queueing model, the processing capability for model updating is characterized by the computing rate $\mu$. To enable the DA-AoM theoretical model to guide the practical deployment and optimization of edge single-server model updating, hardware computing performance is calibrated as $\mu_{0} = ({P_{\text{GPU}} \times \eta})/{C_{base}}$, where $P_{\text{GPU}}$ is the peak computational power of graphics processing unit (GPU), $\eta$ is the actual utilization rate, and the $C_{base}$ is the computational cost of a single update for the baseline model. In actual operation, considering the resource allocation coefficient $\mu_c\in [0,1]$, the effective computational rate is $\mu_{\text{eff}}=\mu_0\times \mu_c$.

Incorporating the computational load $\phi(\alpha)$ characterized in Section~\ref{section4.2}, the average DA-AoM under the M/M/1 queueing model is given by
\begin{equation}
\Delta_{\mathrm{M/M/1}}
(\alpha,\lambda,\mu_{\mathrm{eff}})
= u(\alpha)\Bigg[
\frac{1}{\lambda}
+
\frac{\lambda^{2}\phi^{3}(\alpha)}
{\mu_{\mathrm{eff}}^{2}
\left(\mu_{\mathrm{eff}}-\lambda\phi(\alpha)\right)}
\Bigg].
\label{eq:daaom_mm1}
\end{equation}
This expression clearly illustrates the combined effects of $u(\alpha)$ and $\phi(\alpha)$ on DA-AoM. The model performance decay function $u(\alpha)$ serves as a global weighting coefficient, indicating that stronger data distribution drift produces greater semantic loss under the same delay. In the measured model, $\phi(\alpha)$ remains constant regardless of changes in $\alpha$, and the variation in DA-AoM with respect to $\alpha$ is entirely determined by $u(\alpha)$. In the theoretical model, $\phi(\alpha)$ adapts to changes in $\alpha$, and the variation of DA-AoM is jointly affected by $u(\alpha)$ and $\phi(\alpha)$.

\subsubsection{Cloud DA-AoM With Single Server-D/M/1}

In the D/M/1 queueing model, the update request interval is fixed at $D$, and $\lambda=\frac{1}{D}$. The main factors affecting cloud-side model update DA-AoM include update request generation interval $D$, data distribution drift $\alpha$ and computing rate $\mu$. Using the same hardware calibration method as in the M/M/1 case to calculate the effective computing rate $\mu_{\text{eff}}$, the closed-form expression of the average DA-AoM in the cloud scenario is extended based on the benchmark work as 

\begin{equation}
\Delta_{\mathrm{D/M/1}}
(\alpha,\lambda,\mu_{\mathrm{eff}})
=
u(\alpha)
\left[
\frac{D}{2}
+
\frac{\phi(\alpha)}
{\mu_{\mathrm{eff}}(1-\sigma)}
\right].
\label{eq:daaom_dm1}
\end{equation}
where the congestion factor $\sigma$ reflects the probability that a task finds the server busy upon arrival, and it is the unique real root of the following transcendental equation within the interval $(0,1)$:

\begin{equation}
    \sigma = \exp\left( - \mu_{\text{eff}}D(1 - \sigma)\text{/}\phi \right).
    \label{sigma}
\end{equation}

When the model update frequency is $\lambda\to0$, $\sigma\to 0$, and the queueing time disappears. When $\lambda\to \mu_{\text{eff}}/\phi$, $\sigma\to1$, the denominator of the closed-form expression approaches zero, causing DA-AoM to increase sharply.

\subsubsection{Performance Comparison of Theoretical DA-AoM Guided Model Updates Between Edge and Cloud Layers}

 By comparing the closed-form expressions of M/M/1 and D/M/1, the performance improvement brought by deterministic arrivals can be quantified theoretically. Under the same computational load $\phi$, the average queueing waiting time of the D/M/1 queueing model is lower than that of the M/M/1 queueing model. Although the edge layer (M/M/1) is closer to the device, its random arrival of update requests makes congestion more likely under the same load, so it is suitable for low-frequency and highly responsive tasks. Although the cloud layer (D/M/1) has network round-trip delay, periodic scheduling enables it to process high-frequency update tasks with higher throughput and less congestion.

\section{DA-AoM-Driven Adaptive Optimal Update Strategy Design}
\label{section5}
Based on the closed-form expressions of the average DA-AoM in the edge-layer and cloud-layer scenarios derived in Section~\ref{section4}, this section further studies the model update frequency optimization problem in resource-constrained SemCom systems. The optimization objective of this paper is to maintain model freshness, that is, to minimize the system average DA-AoM by selecting an appropriate model update frequency. Since DA-AoM exhibits a non-monotonic relationship with the model update frequency, an excessively low update frequency leads to insufficient model freshness, whereas an excessively high update frequency causes queueing congestion and increases the update completion time \cite{ref38,ref39}.

This section defines the model update strategy as a DA-AoM-driven adaptive optimal update strategy. Based on the closed-form expressions of DA-AoM obtained in Section~\ref{section4}, the strategy solves for the normalized update frequency that minimizes the average DA-AoM under different queueing models. Specifically, this section analyzes the variation trend of DA-AoM with normalized update frequency and derives the corresponding optimal update frequency for the edge-layer M/M/1 random update scenario and the cloud-layer D/M/1 periodic update scenario, respectively.

\subsection{Normalized Update Frequency Modeling}
\label{section5.1}
In resource-constrained SemCom systems, model update frequency $\lambda$ is a key decision variable affecting DA-AoM. A low update frequency increases model staleness and makes it difficult for the semantic codec model to adapt to target-domain changes in time. A high update frequency can reduce model staleness, but it increases the arrival density of update requests and may cause queueing congestion. Therefore, model update frequency needs to be designed according to the system service capability. 

According to the definitions in Section~\ref{section4}, let $\mu_0$ denote the system baseline computing rate and $\mu_c$ denote the computational resource allocation coefficient. The effective computing rate of the current model update server is then $\mu_{\text{eff}}=\mu_0\times\mu_c$. The data distribution drift intensity changes the relative computational load required by a single model update, and the actual model update frequency is defined as $\lambda$. Under drift intensity $\alpha$, the stability condition of the model update system is $\lambda\phi(\alpha)<\mu_{\text{eff}}$. Therefore, the maximum stable actual update frequency under the current drift state is $\lambda_\text{max}(\alpha) = {\mu_{\text{eff}}}/{\phi(\alpha)}$.

Since $\lambda_{max}(\alpha)$ changes with drift intensity, directly using the actual update frequency $\lambda$ leads to different feasible ranges under different drift states, which is inconvenient for subsequent strategy analysis. Therefore, this paper introduces the normalized update frequency

\begin{equation}
    \widetilde{\lambda} = \frac{\lambda}{\lambda_{\max}(\alpha)} = \frac{\lambda\phi(\alpha)}{\mu_{\text{eff}}},0 < \widetilde{\lambda} < 1.
    \label{eq:norm_lam}
\end{equation}

This transformation maps the drift dependent feasible range of the actual update frequency to a fixed interval $(0,1)$. When $\widetilde{\lambda}$ is small, there are fewer model update requests, and the system has low queueing pressure. When $\widetilde{{\lambda}}$ approaches 1, update requests approach the current service capacity limit, and queueing delay increases rapidly. Therefore, the normalized update frequency can characterize the load level of the model update relative to the current service capacity.

\subsection{DA-AoM-Driven Adaptive Optimal Update Strategy}
\label{section4.2}
To analyze the model update frequency in a unified stable interval, this paper solves the optimal DA-AoM operating point based on the normalized update frequency. The goal of the strategy is to minimize the average DA-AoM under a given queueing model and obtain the normalized frequency that optimizes model freshness. The optimization objective of this section can be expressed as

\begin{equation}
    \widetilde{\lambda}_{\chi}^{*} = \arg{\min_{0 < \widetilde{\lambda} < 1}{\overline{\Delta}_{\text{DA-AoM}}\left( \widetilde{\lambda},\alpha \right)}},\quad\chi \in \text{\{}\text{M},\text{D}\text{\}}.
    \label{eq:argmin}
\end{equation}

According to the DA-AoM expressions in Section~\ref{section4}, after normalized update-frequency modeling, the average DA-AoM under the M/M/1 and D/M/1 queueing models can be decomposed into two parts: a drift-sensitive factor determined by the drift intensity and a queueing cost function determined by the normalized update frequency. For queueing model $\chi\in \text{{M,D}}$, the average DA-AoM can be uniformly written as

\begin{equation}
    {\overline{\Delta}}_{\chi}\left( \widetilde{\lambda},\alpha \right) = S(\alpha)G_{\chi}\left( \widetilde{\lambda} \right),\quad\chi \in \text{\{}\text{M},\text{D}\text{\}},
    \label{eq:chi}
\end{equation}
where $\text{M}$ denotes the M/M/1 queue, $\text{D}$ denotes the D/M/1 queue, $S(\alpha)$ denotes the drift-sensitive factor that characterizes the influence of drift intensity on the DA-AoM value, and $G_\chi(\widetilde{\lambda})$ denotes the normalized-frequency cost function under queueing model $\chi$, which characterizes the relationship between update frequency and queueing delay.

Since $S(\alpha)$ is independent of $\widetilde{\lambda}$, when $S(\alpha)>0$, minimizing $\overline{\Delta}_\chi(\widetilde{\lambda},\alpha)$ is equivalent to minimizing $G_\chi(\widetilde{\lambda})$. Therefore, the DA-AoM-driven optimal normalized frequency is defined as

\begin{equation}
    \widetilde{\lambda}_{\chi}^{*} = \arg{\min_{0 < \widetilde{\lambda} < 1}{G_{\chi}\left( \widetilde{\lambda} \right)}},\quad\chi \in \text{\{}\text{M},\text{D}\text{\}}.
    \label{chi_1}
\end{equation}
This definition indicates that drift intensity mainly changes the numerical value of DA-AoM, but does not change the position of the optimal normalized update frequency. Therefore, $\widetilde{\lambda}_{\chi}^*$ is a fixed optimal normalized update frequency, whereas the actual update frequency remains adaptive to data distribution drift and is given by $\lambda_{\chi}^{*}(\alpha)=\widetilde{\lambda}_{\chi}^{*}\frac{\mu_{\mathrm{eff}}}{\phi(\alpha)}.$

The resulting $\lambda_{\chi}^*(\alpha)$ represents the optimal actual update frequency conditioned on the currently estimated cycle-level drift state. When the target-domain distribution changes between update cycles, $\alpha$ and $\phi(\alpha)$ are re-estimated, and the actual update frequency is accordingly reconfigured. If a substantial distribution change occurs within a single update cycle, however, the previously estimated $\alpha_i$ may no longer accurately represents the current target-domain state, causing $\lambda_{\chi}^*(\alpha_i)$ to temporarily deviate from the optimum corresponding to the new drift state. In rapidly varying environments, the drift state can therefore be re-estimated using a shorter observation window or a change-triggered detection mechanism \cite{ref26,ref28}. The deruved frequency should thus be interpreted as a locally optimal operating point under the cycle-level stationary approximation.

\subsubsection{Optimal Update Frequency in the Edge-Layer M/M/1 Scenario}

Based on the theoretical closed-form expression of the average DA-AoM under the M/M/1 queueing model, substituting the normalized update frequency $\widetilde{\lambda}$ into the expression yields the average DA-AoM under the M/M/1 queue as

\begin{equation}
    {\overline{\Delta}}_{\text{M}}\left( \widetilde{\lambda},\alpha \right) = \frac{u(\alpha)\phi(\alpha)}{\mu_{\text{eff}}}\left( \frac{1}{\widetilde{\lambda}} + \frac{\widetilde{\lambda^{2}}}{1 - \widetilde{\lambda}} + 1 \right).
\label{eq:M}
\end{equation}
For simplicity, the drift-sensitive factor $S(\alpha)$ and the normalized-frequency cost function $G_M(\widetilde{\lambda})$ under the M/M/1 queue are defined as

\begin{equation}
    S(\alpha) = \frac{u(\alpha)\phi(\alpha)}{\mu_{\text{eff}}},
\label{eq:mm1_s}
\end{equation}

\begin{equation}
    G_{\text{M}}\left( \widetilde{\lambda} \right) = \frac{1}{\widetilde{\lambda}} + \frac{\widetilde{\lambda^{2}}}{1 - \widetilde{\lambda}} + 1,
    \label{eq:mm1_g}
\end{equation}
where $G_{\text{M}}(\widetilde{\lambda)}$ consists of three components. The first term $1/\widetilde{\lambda}$ represents the model obsolescence caused by low-frequency updates. The second term $\widetilde{\lambda}^2/(1-\widetilde{\lambda})$ represents the queueing congestion cost resulting from high-frequency updates, and the third term represents the normalized basic service time term.

Since the optimal normalized update frequency under the M/M/1 queue is determined by $G_M(\widetilde{\lambda})$, the first derivative of $G_M(\widetilde{\lambda})$ is first calculated to prove the existence of the optimal update frequency

\begin{equation}
    G_{\text{M}}^{'}\left( \widetilde{\lambda} \right) = - \frac{1}{\widetilde{\lambda^{2}}} + \frac{2\widetilde{\lambda} - \widetilde{\lambda^{2}}}{\left( 1 - \widetilde{\lambda} \right)^{2}}.
\end{equation}
The second derivative is further calculated as 

\begin{equation}
    G_{\text{M}}^{''}\left( \widetilde{\lambda} \right) = \frac{2}{\widetilde{\lambda^{3}}} + \frac{2}{\left( 1 - \widetilde{\lambda} \right)^{3}}.
\end{equation}
Since $0<\widetilde{\lambda}<1$, it follows that $G_{\text{M}}^{''}\left( \widetilde{\lambda} \right)>0$. This indicates that $G_{\text{M}}(\widetilde{\lambda})$ is a strictly convex function within the normalized frequency range. Therefore, there exists a unique normalized update frequency that minimizes the DA-AoM in an M/M/1 queue. Let $G_{\text{M}}^{'}\left( \widetilde{\lambda} \right)=0$, in the idealized approximation analysis, this optimal point can be expressed as $\widetilde{\lambda}_{\text{M}}^*\approx 0.5$. This result indicates that in the edge-layer random update scenario, the optimal update frequency under the M/M/1 queue should lie in the middle region of the maximum stable update capability of the system. Considering the fixed service time term and statistical errors in practical systems, the real optimal point may slightly deviate from 0.5.

\subsubsection{Optimal Update Frequency in the Cloud-Layer D/M/1 Scenario}

Compared with the M/M/1 queue, the D/M/1 queue eliminates the randomness of the request arrival process but is still affected by service-time randomness and queueing waiting. According to the relationship between the actual model update interval $D$ and the normalized frequency, we have $D = {\phi(\alpha)}/{\widetilde{\lambda}\mu_{\text{eff}}}.$

According to the average DA-AoM expression under the D/M/1 queue in Section~\ref{section4}, substituting the above normalized frequency relationship yields

\begin{equation}
    {\overline{\Delta}}_{\text{D}}\left( \widetilde{\lambda},\alpha \right) = \frac{u(\alpha)\phi(\alpha)}{\mu_{\text{eff}}}\left( \frac{1}{2\widetilde{\lambda}} + \frac{1}{1 - \sigma\left( \widetilde{\lambda} \right)} \right).
\end{equation}
For simplicity, the drift-sensitive factor $S(\alpha)$ and the normalized-frequency cost function $G_D(\widetilde{\lambda})$ under the D/M/1 queue are defined as

\begin{equation}
    S(\alpha) = \frac{u(\alpha)\phi(\alpha)}{\mu_{\text{eff}}},
\end{equation}

\begin{equation}
    G_{\text{D}}\left( \widetilde{\lambda} \right) = \frac{1}{2\widetilde{\lambda}} + \frac{1}{1 - \sigma\left( \widetilde{\lambda} \right)}.
\end{equation}

Similarly, the optimal normalized frequency under the D/M/1 queue is determined by $G_D(\widetilde{\lambda})$:

\begin{equation}
    \widetilde{\lambda}_{\text{D}}^{*} = \arg{\min_{0 < \widetilde{\lambda} < 1}{G_{\text{D}}\left( \widetilde{\lambda} \right)}}.
\end{equation}
When $\widetilde{\lambda}\to0$, the term $1/(2\widetilde{\lambda})$ tends to infinity, indicating that the fixed update interval is too long and the model cannot be updated in time. When $\widetilde{\lambda}\to1$, the system approaches the stability boundary, $\sigma(\widetilde{\lambda})$ approaches 1, and the term $1/(1-\sigma(\widetilde{\lambda}))$ increases rapidly, indicating a significant increase in service waiting. Therefore, the ideal optimal frequency under the D/M/1 queue does not appear at the boundary but should lie inside the interval.

Since $\sigma(\tilde{\lambda})$ is implicitly determined by a transcendental equation, an explicit closed-form solution for the optimal normalized frequency in the D/M/1 queue is difficult to obtain. Compared with the M/M/1 case, the staleness term in the D/M/1 expression is reduced from $1/\tilde{\lambda}$ to $1/(2\tilde{\lambda})$, indicating that deterministic update arrivals can reduce the contribution of inter-arrival staleness. However, the final optimal frequency is not determined by the staleness term alone, but also by the congestion term $1/(1-\sigma(\tilde{\lambda}))$, which increases as the system approaches the stability boundary. Therefore, the optimal normalized update frequency for the D/M/1 queue is obtained by numerically minimizing $G_\text{D}(\tilde{\lambda})$ over $0<\tilde{\lambda}<1$.

The optimal normalized frequency is queue-model dependent rather than invariant to the arrival process. Under more general arrivals processes, the DA-AoM formulation in (\ref{eq:average daaom_lam}) can still be used be evaluating the corresponding inter-arrival and system-time statistics, i.e,  $E[X_i^2]$ and $E[X_iQ_i]$. More bursty arrivals may increase queueing delay and shift the optimal operating point toward a more conservative update frequency, while the fundamental tradeoff between model staleness at low update rates and queueing congestion at high update rates remains unchanged. The M/M/1 and D/M/1 results in this work should thus be interpreted as closed-form optimal operating points for two representative arrival models.

\section{Simulation Verification of the DA-AoM Metric and Model Update Strategy}
\label{section6}
This section verifies the DA-AoM theoretical model proposed in Section~\ref{section4} and the model update strategy proposed in Section~\ref{section5} through simulations. Unlike conventional age metrics that only focus on temporal staleness, DA-AoM aims to characterize the freshness of semantic codec models under data distribution drift and further provide a basis for update-frequency selection under resource constraints. Therefore, this section first compares the trend consistency among DA-AoM, AoI, and measured model performance variations to verify the effectiveness of DA-AoM as a freshness metric for semantic codec models. It then analyzes the impacts of key parameters such as data distribution drift intensity $\alpha$, computing rate $\mu$, network bandwidth, and update frequency $\lambda$ on DA-AoM. On this basis, the performance advantage of the DA-AoM-driven optimal update strategy over fixed-frequency, threshold-triggered, and function-based adaptive update strategies is further verified.

This paper uses the DIV2K dataset as the source-domain data and the Places365 dataset as the target-domain data to simulate data distribution drift encountered by SemCom systems during deployment. The Places365 dataset contains 365 scene categories. In the experiment, 80\% of the images in each category are used as the training set, and the remaining 20\% of the images are used as the test set. The semantic codec model is SwinJSCC-Large pretrained on the DIV2K dataset. After target-domain samples are input into the semantic codec model, the drift measurement method proposed in Section~\ref{section3} is used to calculate the data distribution drift intensity $\alpha$, and the model performance degradation rate $u(\alpha)$ is obtained by combining transmission performance metrics. For the computational load $\phi(\alpha)$ part, the theoretical drift-aware load model and the measured fixed load model given in Section~\ref{section4} are used respectively to compare the performance difference between ideal adaptive update capability and the actual fixed network structure. 


The theoretical DA-AoM is first calculated using the closed-form expressions derived in Section~\ref{section4}, where the drift-dependent computational load $\phi(\alpha)$ and performance degradation rate $u(\alpha)$ are determined from the corresponding fitted relationships. To validate the analytical result, we further obtain a measured DA-AoM through Monte Carlo simulation of the model-update process. In the simulation, update arrivals and service processes are generated according to the corresponding queueing model, while the computation-related parameters are calibrated using the deployed SwinJSCC-Large implementation. Since the deployed network is used as the implementation-level service reference rather than introducing an additional computational-load model.

For each parameter configuration, the simulation is independently repeated five times to reduce the influence of stochastic fluctuations. The instantaneous DA-AoM is accumulated over the obsrvation interval, and its time-average value is calculated as
\begin{equation}
    {\overline{\Delta}}_{\text{DA-AoM}} = \frac{1}{T_{total}}\int_{0}^{T_{total}} u\left( \alpha(t) \right) \cdot \left( t - t_{gen}(t) \right)dt.
    \label{eq:measured}
\end{equation}

\subsection{Validity of the DA-AoM Metric and Parameter Impact Analysis}

Taking the M/M/1 edge single-server model update scenario as an example, this section analyzes the impacts of key parameters, including data distribution drift intensity $\alpha$, computing rate $\mu$, network bandwidth, and update frequency $\lambda$, on DA-AoM, and compares the proposed DA-AoM with conventional AoI.

\begin{figure}[!t]
    \centering
    \includegraphics[width=0.9\linewidth]{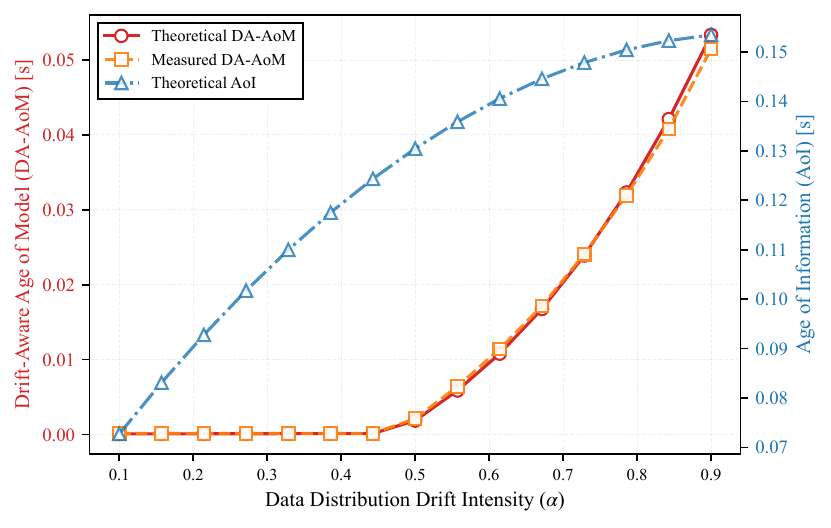}
    \caption{DA-AoM and AoI under different data distribution drift.}
    \label{fig:alpha}
    \vspace{-5mm}
\end{figure}

 \begin{figure}
     \centering
     \includegraphics[width=0.9\linewidth]{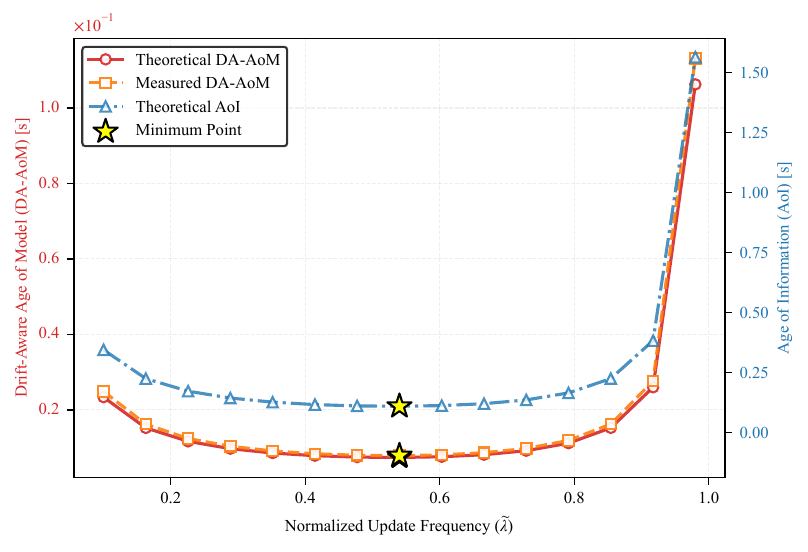}
     \caption{DA-AoM and AoI under different normalized update frequency.}
     \label{fig:lambda}
     \vspace{-5mm}
 \end{figure}
 
Figure~\ref{fig:alpha} shows the average theoretical and measured DA-AoM
and AoI as the drift intensity $\alpha$ increases, with
$\widetilde{\lambda}=0.6$ and $\mu=0.8$. Both theoretical and measured
DA-AoM exhibit nonlinear growth with increasing drift, while AoI varies
more smoothly because it does not incorporate drift-induced performance
degradation.

In the low-drift region, DA-AoM increases slowly due to the generalization
capability of the semantic codec model. As the drift becomes more severe,
DA-AoM increases more rapidly as performance degradation contributes more
strongly to freshness loss. In contrast, AoI cannot distinguish the effects
of mild and severe drift. Therefore, DA-AoM better captures both temporal
staleness and drift-induced model degradation in dynamic SemCom systems.

Figure~\ref{fig:lambda} shows the variation trends of average DA-AoM, average AoI, and measured average DA-AoM as the normalized update frequency increases. Data distribution drift ($\alpha$) is set to 0.6, and the normalized computing rate ($\mu$) is set to 0.8. As shown in the figure, when data distribution drift is fixed, the three metrics show similar trends, and an optimal value can be found at the same update frequency, marked by a star. This indicates that DA-AoM can guide the model to find the best update frequency, which provides an operating point that balances staleness and queueing delay. In addition, DA-AoM remains a more desirable performance metric than AoI, because AoI is insensitive to model performance degradation and may cause network resource waste when guiding model updates.

\begin{figure}
    \centering
    \includegraphics[width=0.9\linewidth]{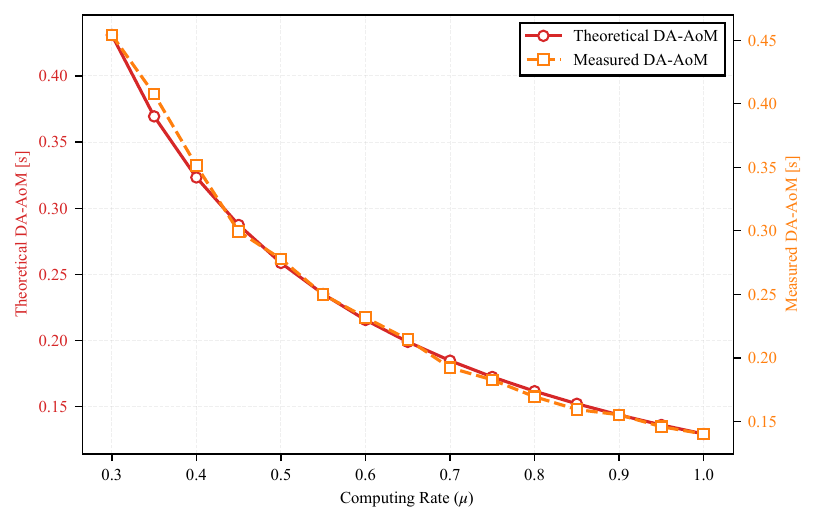}
    \caption{ DA-AoM under different computing rate.}
    \label{fig:mu}
    \vspace{-5mm}
\end{figure}

\begin{figure}
    \centering
    \includegraphics[width=0.9\linewidth]{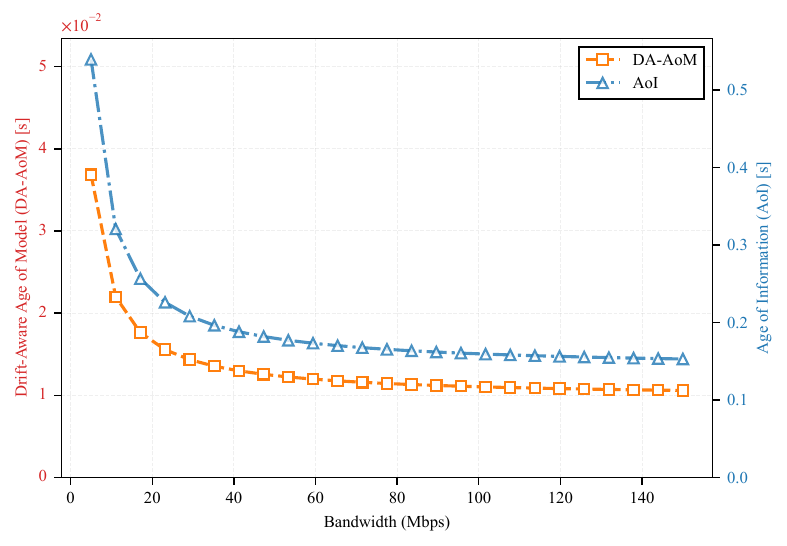}
    \caption{DA-AoM and AoI under different bandwidth.}
    \label{fig:bandwidth}
    \vspace{-3mm}
\end{figure}

Figure~\ref{fig:mu} shows the variation trends of average DA-AoM and measured average DA-AoM as the computing rate ($\mu$) increases. As shown in the figure, both curves continuously decrease as the computing rate increases. Over the entire computing rate range, both theoretical and measured results show a rapid decline in the low-rate region and a gradually slowing decline in the high-rate region. The trend consistency between measured DA-AoM and theoretical DA-AoM verifies that the proposed DA-AoM metric remains effective under an actual fixed model structure.

Figure~\ref{fig:bandwidth} shows the variation trends of average DA-AoM and average AoI as bandwidth increases. Data distribution drift ($\alpha$) is set to 0.6, and the normalized update frequency ($\widetilde{\lambda}$) is set to 0.6. As shown in the figure, the two metrics show similar decreasing trends. This is because the bandwidth mainly affects the network transmission delay $T_{\mathrm{net}}=D_{\text{model}}/R_{\text{channel}}$, which is outside the computation queueing system. Increasing the bandwidth reduces $T_{\text{net}}$, thereby decreasing the end-to-end completion delay of model updates and reducing the corresponding freshness loss.

The transmission delay $T_{\text{net}}$ is treated as an external delay term and does not change the queueing stability condition or the optimal normalized update frequency derived from the computation side queueing model. Therefore, the bandwidth experiment is used to evaluate the sensitivity of DA-AoM to communication delay, rather than to change the optimal frequency analysis. Compared with AoI, DA-AoM also includes the model performance degradation rate caused by data distribution drift, and thus reflects the effects of both communication delay and drift intensity on model freshness.

\subsection{Numerical Structure Analysis of DA-AoM}

\begin{figure}[!t]
    \centering    \includegraphics[width=0.9\linewidth]{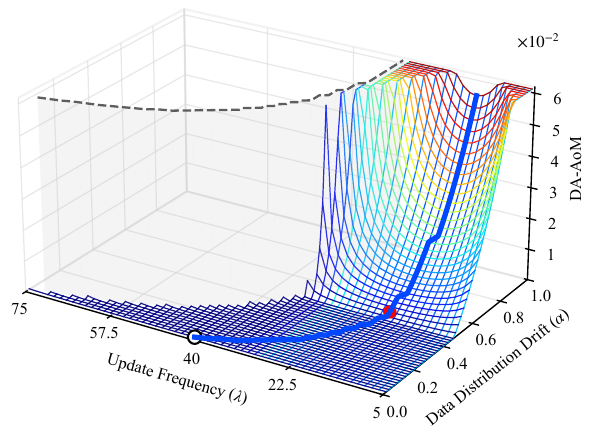}
    \caption{Impact of data distribution drift and update frequency on DA-AoM for edge-layer model updating (M/M/1).}
    \label{fig:mm1_3D}
    \vspace{-5mm}
\end{figure}

\begin{figure}[!t]
    \centering
    \includegraphics[width=0.9\linewidth]{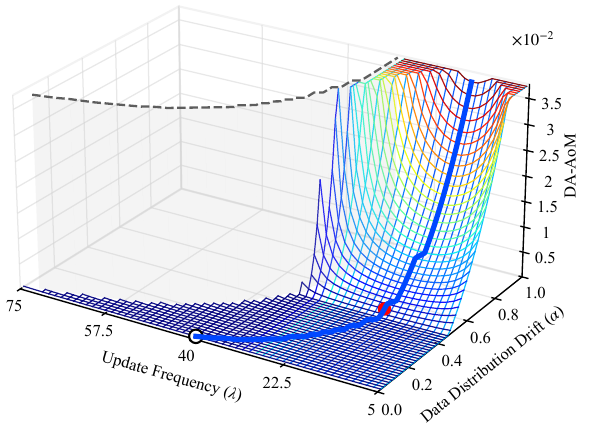}
    \caption{Impact of data distribution drift and update frequency on DA-AoM for cloud-layer model updating (D/M/1)}
    \label{fig:dm1_3D}
    \vspace{-5mm}
\end{figure}

Figures~\ref{fig:mm1_3D} and~\ref{fig:dm1_3D} show the three-dimensional surfaces of average DA-AoM with respect to model update frequency $\lambda$ and data distribution drift intensity $\alpha$ under the edge M/M/1 queue and cloud D/M/1 queue, respectively. The computing rate $\mu$ is set to 0.8. The blue trajectory represents the optimal update frequency trajectory that minimizes DA-AoM under the current drift intensity. The hollow circle represents the global minimum DA-AoM point. The red point represents the point on the optimal trajectory where DA-AoM starts to become significantly sensitive to drift changes. The gray dashed line represents the boundary of the maximum stable update frequency determined by $\lambda_{\text{max}}(\alpha)=\mu_{\text{eff}}/\phi(\alpha)$, and the gray shaded region represents the unstable queueing update region beyond this boundary.

From these two figures, it can be seen that DA-AoM first decreases and then increases with the update frequency. When the update frequency is low, model updating is not timely and DA-AoM is large. When the update frequency is increased appropriately, the model can complete updates more quickly and DA-AoM decreases. However, when the update frequency continues to increase and approaches the gray stability boundary, queueing waiting time increases rapidly and DA-AoM rises again. The gray boundary in the figure further indicates that the model update frequency is constrained by the queueing load. The system must satisfy $\lambda<\lambda_\text{max}(\alpha)$. Once the update frequency exceeds this boundary, the arrival rate of update requests will exceed the service capacity of the compute nodes, leading to persistent queue backlog and causing the system to enter an unstable region. Therefore, the model update strategy cannot rely solely on increasing the update frequency to reduce DA-AoM, it must also select the optimal update frequency within the range that the load factor can tolerate.

Meanwhile, as data distribution drift intensity $\alpha$ increases, the overall DA-AoM surface rises, indicating that stronger drift causes more severe model performance degradation and greater model freshness loss under the same update frequency. The blue optimal trajectory also shows that the optimal update frequency $\lambda$ corresponding to different drift intensities is different, and the system needs to select an appropriate operating point according to drift intensity and computing service capability. 

Comparing Figures~\ref{fig:mm1_3D} and~\ref{fig:dm1_3D}, the surface in the cloud D/M/1 scenario is smoother overall, and the increase of DA-AoM in the high-frequency region is relatively milder. In the M/M/1 scenario, because update requests arrive randomly, queueing congestion is more likely to occur in the high-frequency region. Therefore, compared with the edge update scenario with random arrivals, the cloud update scenario with fixed-interval arrivals has more stable update performance.

\subsection{Verification of the DA-AoM-Driven Adaptive Optimal Update Strategy}

\begin{figure}[!t]
    \centering
    \includegraphics[width=0.9\linewidth]{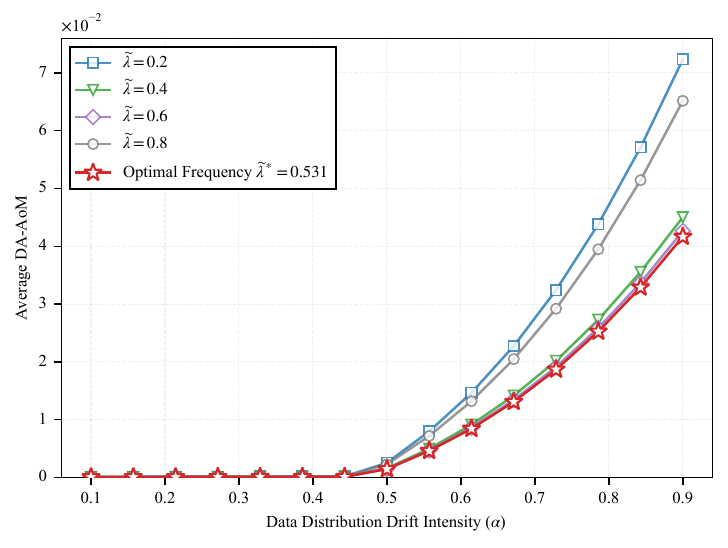}
    \caption{Optimal update frequency of the DA-AoM-driven model update strategy at the edge layer (M/M/1)}
    \label{fig:mm1_optimal}
\end{figure}

\begin{figure}[!t]
    \centering
    \includegraphics[width=0.9\linewidth]{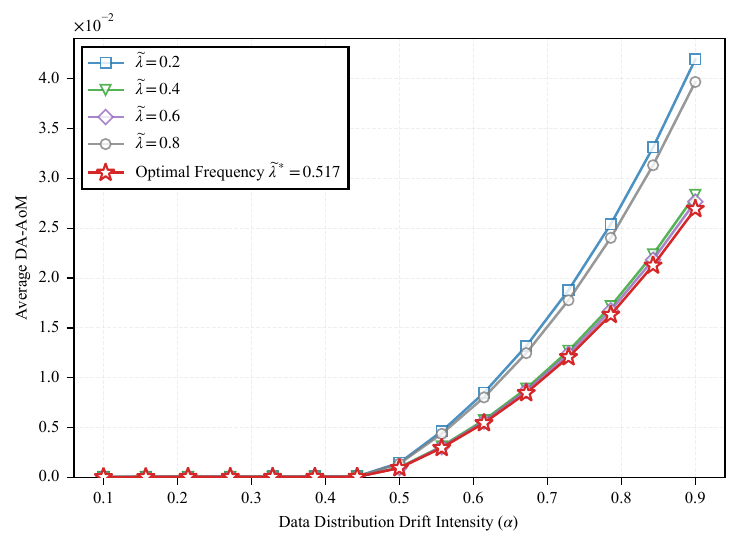}
    \caption{Optimal update frequency of the DA-AoM-driven model update strategy at the cloud layer (D/M/1)}
    \label{fig:dm1_optimal}
\end{figure}

Figures~\ref{fig:mm1_optimal} and~\ref{fig:dm1_optimal} show the performance of the DA-AoM-driven adaptive optimal update strategy proposed in Section~\ref{section5} at the edge layer (M/M/1) and cloud layer (D/M/1), respectively. To verify the effectiveness of the adaptive optimal normalized update frequency, the fixed optimal normalized update frequency derived in the figure is compared with multiple fixed update frequency strategies. As shown in the figure, whether at the edge layer or the cloud layer, the DA-AoM-driven optimal strategy, marked by the red star, maintains a lower average DA-AoM under different data distribution drift intensities. This indicates that DA-AoM can provide an effective frequency selection basis for model updating. By minimizing the average DA-AoM, the system can determine a globally stable normalized update frequency, thereby avoiding model aging caused by fixed low-frequency updating and avoiding increased queueing delay caused by fixed high-frequency updating.

Specifically, the optimal normalized update frequency for the edge-layer M/M/1 scenario is $\widetilde{\lambda}_{\text{M}}^*=0.531$, and for the cloud-layer D/M/1 scenario, it is $\widetilde{\lambda}_{\text{D}}^*=0.517$. These results validate the existence and robustness of the optimal normalized update frequency, demonstrating that DA-AoM can be used to guide the system in adaptively selecting a reasonable model maintenance frequency.


\subsection{Comparative Analysis of Update Strategy Performance}

To verify the effectiveness of the DA-AoM-driven adaptive update strategy proposed in Section~\ref{section5}, this section compares the proposed strategy with several typical update strategies in the M/M/1 edge-layer model update scenario. Specifically, this section not only compares the DA-AoM values corresponding to the update frequencies of different strategies, but also further analyzes the performance gain brought by the proposed strategy.

To reflect the impact of different update strategies on model maintenance performance, this paper selects the following five types of baseline strategies for comparison:

(1) Proportional update strategy ($\lambda \propto \alpha$): this strategy increases the update frequency linearly with data distribution drift intensity and is used to simulate a simple drift-adaptive updating method.

(2) Threshold-triggered update strategy: this strategy switches to a higher update frequency after the drift intensity exceeds a preset threshold, that is, when $\alpha>0.5$, higher-frequency model updating is triggered. 

(3) Square-root update strategy ($\lambda\propto\sqrt{\alpha}$): this strategy makes the update frequency increase sublinearly with drift intensity, and its frequency variation depends on a preset function. 

(4) Fixed low-frequency update strategy ($\widetilde{\lambda}=0.2$): a conservative strategy that aims to minimize queueing delay, but faces severe semantic obsolescence risk because of its low update frequency.

(5) Fixed high-frequency update strategy ($\widetilde{\lambda} = 0.8$): an aggressive strategy that aims to reduce the sampling interval through high-frequency updating, but it can easily cause queueing congestion.

\begin{figure}[!t]
    \centering
    \includegraphics[width=0.9\linewidth]{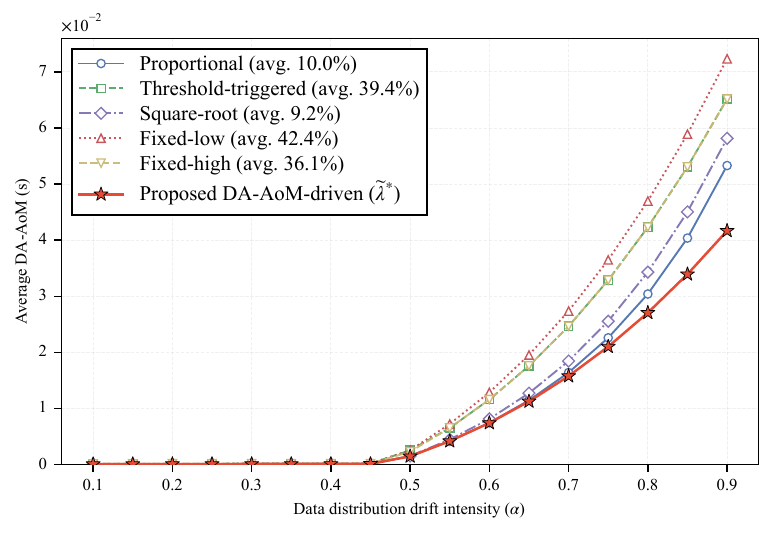}
    \caption{Comparison between the DA-AoM-driven optimal model update strategy and baseline update strategies.}
    \label{fig:5baseline}
    \vspace{-5mm}
\end{figure}

Figure~\ref{fig:5baseline} compares the DA-AoM achieved by different
update strategies under varying data distribution drift intensities.
As the drift intensity increases, the DA-AoM of all strategies
increases, while the proposed strategy consistently maintains the
lowest value. The fixed high-frequency strategy suffers from increased
queueing delay, whereas the fixed low-frequency strategy leads to
model staleness due to insufficient updating. The square-root and
threshold-triggered strategies provide greater adaptability but remain
suboptimal because they cannot jointly characterize drift-induced
performance degradation and update queueing delay. On average, the
proposed strategy reduces DA-AoM by 10.0\%, 39.4\%, 9.2\%, 42.4\%,
and 36.1\% compared with the proportional, threshold-triggered,
square-root, fixed low-frequency, and fixed high-frequency strategies,
respectively. These results demonstrate that the proposed strategy can
select a more appropriate update frequency under the queueing stability
constraint and maintain better model freshness under the same resource
constraints.

\section{Conclusion}
\label{section7}
This paper proposes a drift-aware adaptive semantic codec update strategy for dynamic SemCom and introduces DA-AoM as a model freshness metric. DA-AoM jointly captures drift-induced performance degradation and drift-dependent update computational load. Closed-form average DA-AoM expressions are derived for edge-layer M/M/1 random updates and cloud-layer D/M/1 periodic updates, based on which the update frequency is optimized under the queueing-stability constraint. Simulation results show that DA-AoM more effectively characterizes the effects of drift intensity, computing rate, bandwidth, and update frequency on model freshness than conventional AoI. The proposed strategy consistently achieves lower average DA-AoM than representative baseline strategies under different drift conditions, validating DA-AoM for model freshness evaluation and adaptive update optimization. The formulation can be extended to other semantic tasks using task-appropriate drift representations and performance metrics.
\section*{Acknowledgments}
This work was supported in part by the National Natural  Science Foundation of China (Grant No. 62501067 and U24B20131).

\bibliographystyle{IEEEtran}
\bibliography{TCCN_references}


\end{document}